\documentstyle[preprint,aps]{revtex}

\begin{document}

\draft

\title{Extraction of the $\pi NN$ coupling constant \\
       from NN scattering data}
\author{Richard A. Arndt, Igor I. Strakovsky$^\dagger$ and Ron L. Workman}
\address{Department of Physics, Virginia Polytechnic Institute and State
University, Blacksburg, VA 24061}

\date{\today}
\maketitle

\begin{abstract}
We reexamine Chew's method for extracting the $\pi NN$ coupling constant
from np differential cross section measurements. Values for this coupling
are extracted below 350 MeV, in the potential model region, and up to
1 GeV. The analyses to 1~GeV have utilized 55 data sets.
We compare these results to those obtained via $\chi^2$ mapping
techniques. We find that these two methods give consistent results which
are in agreement with previous Nijmegen determinations.

\end{abstract}
\vspace*{0.5in}
\pacs{PACS numbers: 13.75.Cs, 13.75.Gx, 21.30.+y}

\narrowtext
\section{Introduction}
\label{sec:intro}
While the $\pi NN$ coupling constant has been known within a few percent
for decades, there have been numerous attempts to more precisely
determine its value, and to search for possible charge-independence
breaking. Recent results from the Nijmegen
$NN$ and $\bar N N$ analyses\cite{Nijmegen},
and the Virginia Tech $\pi N$ and $NN$ analyses\cite{VPI1},\cite{VPI2}
have found values of $g^2/4\pi$ near 13.7 with little evidence for charge
dependence. However, the importance of this coupling continues to motivate
independent studies\cite{other} utilizing different methods of extraction.

In 1958, Chew suggested\cite{Chew}
that the influence of the pion-pole might be
measureable in $NN$ elastic scattering, if appropriate kinematic
regions were examined. This technique has recently been applied to
$\bar p p \to \bar n n$\cite{Bradamante} and
np elastic\cite{no93} differential
cross section measurements.  In these studies, the extracted
values for $g^2/4\pi$ ($g^2/4\pi = 12.83 \pm 0.36$\cite{Bradamante} and
$g^2/4\pi = 12.47 \pm 0.36$\cite{no93}) were somewhat below
those found in Refs.\cite{Nijmegen},\cite{VPI2}.

In the present work, selected np differential cross section
data\cite{no93}$-$\cite{ja84} below 1~GeV have been
analyzed using a modified form of the Chew method\cite{Chew},\cite{er95p}.
In the region below 350 MeV, the ``difference
method"\cite{er95p} was used in conjunction with the Nijmegen,
Bonn and Paris potentials, and various partial-wave solutions. We considered
differences between ``model" and ``experimental" values of the dimensionless
quantity $y \; = \; x^2 \; s\;(d\sigma/ d\Omega)/(\hbar c)^2$ expanded in
terms of a function of $x\; =\; 1 + MT(1 + \cos \theta )/\mu^2$,
\begin{equation}
y_{\rm fit} \; = \; y_{\rm model} \; + \; \sum_{j=0}^{N} a_j b_j(x)
\end{equation}
where $s$ is the usual Mandelstam variable, $T$ is the laboratory kinetic
energy, and $\mu$ ($M$) is the
pion (nucleon) mass. While it is conventional to expand
($y_{\rm fit} \; - \; y_{\rm model}$) as a power series in $x$,
we chose to expand in terms of Legendre polynomials, $P_j(\rho)$,
\begin{equation}
b_0\; =\; 1, \;\;\; b_j(j > 0) \; =\; P_j(\rho) - 1,
\end{equation}
with $\rho \; = 1 \; - 2 (x/x_{\rm max})$,
which resulted in improved numerics
but had no effect on the final results (several different expansions were
made for comparison purposes).
The difference between the model coupling constant and the experimental
value is then given in terms of the first coefficient
\begin{equation}
g^4_{\rm exp} \; = \; g^4_{\rm model} \; + \; a_0,
\end{equation}
in the above expansion.

The results of our analyses are given in Tables~I and II. In Table~I, we
list values of $g^2/4\pi$ found via the difference method applied
to data below 1~GeV,
using our partial-wave solution SM95 as input. In order to treat the
different datasets as consistently as possible, we put a number of
constraints on the fits.
\begin{itemize}
\item
Extractions were performed on data backward of 110 degrees.
\item
Only those datasets with more than 15 qualifying points were accepted.
\item
An upper limit for the variable $x$ was taken to be 10.
\item
The number of fitted parameters was given a lower limit of 4.
\end{itemize}
The search was terminated either at a minimum $\chi^2$/(degree of freedom)
or when the error on $g^2/4\pi$ exceeded 1.

In order to gauge the effect of renormalizing datasets, two independent
determinations of $g^2/4\pi$ were made for each experiment. In the first
case, data were fitted without any renormalization. In the second
case, data were first renormalized according to the input solution SM95.
The average values of $g^2/4\pi$, and their associated uncertainties,
are listed in Table~II for a number of
input solutions and energy ranges. The uncertainties were calculated
without accounting for systematic errors. We estimate that more
realistic uncertainties are larger by at least a factor of 2.

The values in Table I show considerable scatter, as is evident from
Fig.~1. Given this variability, the calculated averages are
remarkably consistent. Coupling constants determined using
potential model and partial-wave inputs below 400 MeV are in
good agreement those found using our partial-wave solutions up
to 1 GeV. These averages are also quite consistent with the results of
Refs.\cite{Nijmegen}, \cite{VPI2}.

As a check, we have repeated the coupling constant extractions of
Refs.\cite{no93}, \cite{er95p}
and agree with their results. In Ref.\cite{er95p} it is claimed that the
96~MeV and 162~MeV Uppsala cross
sections\cite{ro92}, \cite{er95p} imply a
consistent value for $g^2 /4\pi$. This is not evident from the values
quoted in Table~I. We can, however, explain the discrepancy. While order 4
and order 6 fits to the data of Ref.\cite{ro92} gave values of $g^2/4\pi$
consistent with those found in Ref.\cite{er95p}, the order 5 fit gave a
somewhat larger value. In choosing the results listed in Table~I,
the order 6 fit was
rejected, as the associated error was too large, according to our criterea
for terminating a search. In contrast, the coupling derived from the data of
Ref.\cite{er95p} was much less sensitive to the chosen order.

As a further consistency check we mapped $g^2/4\pi$ against $\chi^2$
over different energy regions, analyzing $pp$ and $np$ data both separately
and in combined fits. This extended a study reported earlier\cite{VPI1}.
In analyzing the combined dataset (pp and np) below 400 MeV, we found a
value of coupling near 13.8. Results were obtained using the Nijmegen
Coulomb distorted-wave Born approximation\cite{niem} for the one-pion
exchange (OPE) contribution.  A significantly higher value of $g^2/4\pi$
(near 14.4) was found if we used a plane-wave Born
approximation\cite{ar83} for the OPE contribution. While this appears
to confirm our Chew-extrapolation results, less consistency was found
when the extraction was broken into separate pp and np contributions, or
when the energy range was increased to 1.6~GeV. In fact the sign of
($g^2_{\rm pp} - g^2_{\rm np}$) changed when the analysis was
extended from 400~MeV to 1.6~GeV. These results are listed in Table~III.
Results of our analyses to 1.6~GeV are displayed in Fig.~2. The $\chi^2$
map from our analyses of elastic pion-nucleon scattering data is included
for comparison.

In summary, we have found that the value of $g^2/4\pi$ extracted from
NN scattering data is quite consistent with our previous determinations.
The Nijmegen group has also generated $\chi^2$ maps\cite{Nijmegen}
of the Bonner et al.\cite{bo78} data below 350~MeV, obtaining results in
qualitative agreement with ours. However, given the variation of
$g^2/4\pi$ values displayed in Fig.~1, we cannot claim to have improved upon
the results found from our application of dispersion relations to
elastic pion-nucleon scattering data.

\acknowledgments

We thank M. Rentmeester for useful communications, and B. Loiseau and
N. Olssen for providing experimental data prior to publication.
I.S. acknowledges the hospitality extended by the Physics Department of
Virginia Tech.
This work was supported in part by the U.S.~Department of Energy Grant
DE--FG05--88ER40454.

\eject

\eject
%
\newpage
{\Large\bf Figure captions}\\
\newcounter{fig}
\begin{list}{Figure \arabic{fig}.}
{\usecounter{fig}\setlength{\rightmargin}{\leftmargin}}
\item
{Plot of $g^2/4\pi$ determinations from $np$ differential cross sections
versus the laboratory kinetic energy. The input solution was SM95. The
solid line gives the average value. Results for unnormalized data are
displayed.}
\item
{$\chi^2$ maps from fits to all NN data below 1.6 GeV (dot-dashed line),
pp data only (dashed line), and np data only (dotted line). The
result from fits to elastic pion-nucleon scattering data (solid line) is
display for comparison. The variation of $\chi^2$ from minimal values
($\delta \chi^2$) is plotted against $g^2/4\pi$.}
\end{list}
\vfil
\eject
Table~I.
Values of $g^2/4\pi$ extracted using the difference method applied to
data below 1~GeV at backward angles (see text).
The laboratory kinetic energy (T), the
angular range ($\theta$) and number (N) of data used in the analysis,
the order
(M) of the fit, and the corresponding $\chi^2$/(degree of freedom) are
given for each dataset. In the first column of $g^2/4\pi$ values, datasets
were fit without renormalization. In the next column, a normalization
factor (Norm), determined from solution SM95, was applied prior to fitting.
\vskip 10pt
\centerline{
\vbox{\offinterlineskip
\hrule
\hrule
\halign{\hfill#\hfill&\qquad\hfill#\hfill&\qquad\hfill#\hfill&
\qquad\hfill#\hfill&\qquad\hfill#\hfill&\qquad\hfill#\hfill&
\qquad\hfill#\hfill&\qquad\hfill#\hfill&\qquad\hfill#\hfill\cr
\noalign{\vskip 6pt}
T&$\theta$&N&Ref&\hbox{\hskip .1cm}&\hbox{\hskip .1cm}&
(unnormalized)&\hbox{\hskip .1cm}&(normalized)\cr
(MeV)&($^{\circ}$)&\hbox{\hskip .1cm}&\hbox{\hskip .1cm}&M&
$\chi^2/dof$&$g^2/4\pi$&Norm&$g^2/4\pi$\cr
\noalign{\vskip 6pt}
\noalign{\hrule}
\noalign{\vskip 6pt}
 91.0 & $119-177$ & 16 &\cite{st541}& 4 & 0.81 & 12.59~(1.41) & 1.056
                                               & 10.96~(2.36) \cr
 96.0 & $117-179$ & 32 &\cite{ro92} & 4 & 0.61 & 16.50~(0.57) & 0.982
                                               & 16.37~(0.56) \cr
162.0 & $122-178$ & 43 &\cite{bo78} & 4 & 1.51 & 12.96~(0.75) & 1.050
                                               & 13.19~(0.77) \cr
162.0 & $119-179$ & 31 &\cite{er95p}& 4 & 1.05 & 14.65~(0.36) & 0.938
                                               & 14.32~(0.34) \cr
177.9 & $122-179$ & 44 &\cite{bo78} & 4 & 1.08 & 13.06~(0.67) & 1.042
                                               & 13.25~(0.69) \cr
194.5 & $121-179$ & 42 &\cite{bo78} & 4 & 1.54 & 12.15~(0.63) & 1.040
                                               & 12.29~(0.65) \cr
211.5 & $120-178$ & 43 &\cite{bo78} & 4 & 0.76 & 12.75~(0.54) & 1.029
                                               & 12.86~(0.55) \cr
212.0 & $111-177$ & 30 &\cite{ke82} & 6 & 0.69 & 14.67~(0.68) & 0.970
                                               & 13.88~(0.49) \cr
229.1 & $120-178$ & 49 &\cite{bo78} & 4 & 1.36 & 12.80~(0.47) & 1.027
                                               & 12.90~(0.48) \cr
247.2 & $119-178$ & 53 &\cite{bo78} & 4 & 0.73 & 13.21~(0.42) & 1.014
                                               & 13.26~(0.42) \cr
265.8 & $118-179$ & 63 &\cite{bo78} & 4 & 0.84 & 12.97~(0.37) & 1.002
                                               & 12.98~(0.37) \cr
284.8 & $117-179$ & 73 &\cite{bo78} & 4 & 1.14 & 13.32~(0.23) & 1.024
                                               & 13.41~(0.24) \cr
304.2 & $116-179$ & 79 &\cite{bo78} & 6 & 0.97 & 13.16~(0.76) & 0.975
                                               & 12.72~(0.88) \cr
319.2 & $117-177$ & 40 &\cite{ke82} & 6 & 0.85 & 11.85~(0.69) & 0.989
                                               & 11.75~(0.69) \cr
324.1 & $115-179$ & 80 &\cite{bo78} & 4 & 1.10 & 13.57~(0.18) & 1.028
                                               & 13.67~(0.19) \cr
344.3 & $118-179$ & 74 &\cite{bo78} & 4 & 1.12 & 13.14~(0.21) & 1.003
                                               & 13.15~(0.21) \cr
365.0 & $120-179$ & 62 &\cite{bo78} & 5 & 1.33 & 12.72~(0.46) & 1.046
                                               & 12.89~(0.50) \cr
386.0 & $121-179$ & 55 &\cite{bo78} & 4 & 1.44 & 13.51~(0.20) & 1.066
                                               & 13.73~(0.21) \cr
\noalign{\vskip 10pt}}
\hrule}}
\vfill
\eject
Table~I (continued).
\vskip 10pt
\centerline{
\vbox{\offinterlineskip
\hrule
\hrule
\halign{\hfill#\hfill&\qquad\hfill#\hfill&\qquad\hfill#\hfill&
\qquad\hfill#\hfill&\qquad\hfill#\hfill&\qquad\hfill#\hfill&
\qquad\hfill#\hfill&\qquad\hfill#\hfill&\qquad\hfill#\hfill\cr
\noalign{\vskip 6pt}
T&$\theta$&N&Ref&\hbox{\hskip .1cm}&\hbox{\hskip .1cm}&
(unnormalized)&\hbox{\hskip .1cm}&(normalized)\cr
(MeV)&($^{\circ}$)&\hbox{\hskip .1cm}&\hbox{\hskip .1cm}&M&$\chi^2/dof$&
$g^2/4\pi$&Norm&$g^2/4\pi$\cr
\noalign{\vskip 6pt}
\noalign{\hrule}
\noalign{\vskip 6pt}
407.3 & $123-179$ & 52 &\cite{bo78} & 4 & 0.52 & 13.34~(0.20) & 1.065
                                               & 13.56~(0.20) \cr
418.0 & $128-177$ & 20 &\cite{ke82} & 5 & 0.63 & 12.72~(0.52) & 1.010
                                               & 12.70~(0.52) \cr
418.0 & $127-178$ & 28 &\cite{ke82} & 5 & 0.74 & 12.44~(0.49) & 0.986
                                               & 12.37~(0.49) \cr
421.4 & $152-180$ & 42 &\cite{bi75} & 4 & 1.29 & 15.87~(1.31) & 0.949
                                               & 15.51~(1.27) \cr
428.9 & $134-180$ & 51 &\cite{bo78} & 4 & 0.87 & 13.38~(0.26) & 1.006
                                               & 13.41~(0.26)  \cr
450.9 & $135-180$ & 50 &\cite{bo78} & 4 & 0.75 & 12.92~(0.24) & 1.061
                                               & 13.66~(0.56) \cr
457.2 & $152-180$ & 42 &\cite{bi75} & 4 & 1.47 & 14.35~(0.89) & 1.021
                                               & 14.50~(0.90) \cr
459.3 & $127-180$ & 76 &\cite{no93} & 6 & 1.62 & 12.19~(0.54) & 1.112
                                               & 12.68~(0.59) \cr
473.2 & $134-180$ & 54 &\cite{bo78} & 5 & 0.82 & 12.91~(0.46) & 1.073
                                               & 13.33~(0.48) \cr
493.0 & $128-172$ & 20 &\cite{ke82} & 4 & 1.23 & 13.90~(0.46) & 1.003
                                               & 13.91~(0.47) \cr
494.6 & $151-180$ & 43 &\cite{bi75} & 4 & 2.26 & 13.75~(0.74) & 1.069
                                               & 14.16~(0.77) \cr
495.7 & $133-180$ & 62 &\cite{bo78} & 4 & 1.06 & 13.52~(0.16) & 1.053
                                               & 13.74~(0.16) \cr
518.5 & $132-180$ & 67 &\cite{bo78} & 4 & 0.85 & 13.68~(0.14) & 1.047
                                               & 13.86~(0.14) \cr
532.0 & $133-180$ & 70 &\cite{bi75} & 4 & 1.84 & 12.67~(0.26) & 1.045
                                               & 12.82~(0.27) \cr
541.6 & $131-180$ & 71 &\cite{bo78} & 4 & 1.02 & 13.72~(0.14) & 1.038
                                               & 13.87~(0.14) \cr
564.9 & $133-180$ & 69 &\cite{bo78} & 4 & 1.03 & 13.95~(0.15) & 1.019
                                               & 14.03~(0.15) \cr
570.9 & $134-180$ & 69 &\cite{bi75} & 5 & 1.20 & 12.62~(0.41) & 1.086
                                               & 13.36~(0.20) \cr
588.4 & $134-180$ & 67 &\cite{bo78} & 4 & 0.91 & 13.90~(0.17) & 1.021
                                               & 13.98~(0.17) \cr
610.3 & $134-180$ & 68 &\cite{bi75} & 5 & 1.40 & 12.44~(0.44) & 1.056
                                               & 12.68~(0.46) \cr
612.2 & $134-179$ & 67 &\cite{bo78} & 6 & 0.95 & 15.23~(0.58) & 1.019
                                               & 15.19~(0.56) \cr
636.2 & $137-179$ & 65 &\cite{bo78} & 4 & 0.83 & 14.12~(0.23) & 0.980
                                               & 14.03~(0.23) \cr
647.5 & $136-180$ & 47 &\cite{ev82} & 5 & 0.92 & 13.61~(0.40) & 0.993
                                               & 13.62~(0.38) \cr
649.6 & $146-180$ & 50 &\cite{bi75} & 4 & 2.21 & 12.01~(0.51) & 1.043
                                               & 12.18~(0.53) \cr
660.4 & $136-179$ & 66 &\cite{bo78} & 4 & 1.27 & 14.33~(0.22) & 1.001
                                               & 14.34~(0.22) \cr
684.8 & $137-179$ & 65 &\cite{bo78} & 6 & 0.84 & 15.42~(0.70) & 1.016
                                               & 16.19~(0.76) \cr
690.2 & $150-180$ & 42 &\cite{bi75} & 4 & 1.03 & 12.28~(0.50) & 1.044
                                               & 12.47~(0.51) \cr
709.3 & $138-179$ & 63 &\cite{bo78} & 6 & 0.87 & 13.82~(0.89) & 0.997
                                               & 13.99~(0.22) \cr
731.3 & $138-180$ & 57 &\cite{bi75} & 5 & 1.28 & 11.77~(0.52) & 1.061
                                               & 12.49~(0.48) \cr
734.1 & $138-179$ & 62 &\cite{bo78} & 6 & 1.01 & 13.17~(0.75) & 1.006
                                               & 13.30~(0.73) \cr
770.6 & $139-179$ & 63 &\cite{bo78} & 4 & 1.50 & 13.84~(0.12) & 1.068
                                               & 14.12~(0.12) \cr
772.9 & $139-180$ & 58 &\cite{bi75} & 4 & 1.15 & 13.00~(0.20) & 1.067
                                               & 13.22~(0.21) \cr
801.9 & $141-180$ & 50 &\cite{ja84} & 4 & 0.98 & 13.47~(0.19) & 1.091
                                               & 13.81~(0.20) \cr
814.9 & $141-180$ & 56 &\cite{bi75} & 5 & 1.43 & 11.63~(0.47) & 1.053
                                               & 11.89~(0.48) \cr
856.8 & $142-180$ & 54 &\cite{bi75} & 4 & 0.97 & 12.88~(0.25) & 1.054
                                               & 13.04~(0.26) \cr
899.3 & $143-180$ & 52 &\cite{bi75} & 4 & 1.85 & 13.38~(0.25) & 1.098
                                               & 13.69~(0.27) \cr
942.5 & $144-180$ & 50 &\cite{bi75} & 4 & 1.45 & 13.87~(0.27) & 1.056
                                               & 14.07~(0.28) \cr
986.0 & $144-180$ & 49 &\cite{bi75} & 5 & 1.14 & 14.53~(0.42) & 1.112
                                               & 15.20~(0.45) \cr
\noalign{\vskip 10pt}}
\hrule}}
\vfill
\eject
Table~II. Table of average $g^2/4\pi$ values obtained using
partial-wave analyses (SM95, VV40, and NI93),
and potential models (NY93, BONN, and PARIS) as input.
N is the number of datasets involved in the analysis.
\vskip 10pt
\centerline{
\vbox{\offinterlineskip
\hrule
\hrule
\halign{\hfill#\hfill&\qquad\hfill#\hfill&\qquad\hfill#\hfill&
\qquad\hfill#\hfill&\qquad\hfill#\hfill&\qquad\hfill#\hfill&
\qquad\hfill#\hfill\cr
\noalign{\vskip 6pt}
Solution & (Model) & Ref & Range & N & (unnormalized) & (normalized)\cr
\hbox{\hskip .1cm}&$g^2/4\pi$&\hbox{\hskip .1cm}&(MeV)&\hbox{\hskip .1cm}&
$g^2/4\pi$&$g^2/4\pi$\cr
\noalign{\vskip 6pt}
\noalign{\hrule}
\noalign{\vskip 6pt}
SM95 &13.75& present   & $0-1000$ & 55 & 13.58~(0.04) & 13.75~(0.04) \cr
SM95 &13.75& present   & $0-400$  & 18 & 13.47~(0.08) & 13.51~(0.08) \cr
VV40 &13.75& present   & $0-400$  & 18 & 13.55~(0.08) & 13.66~(0.08) \cr
NI93 &13.58&\cite{ni93}& $0-350$  & 16 & 13.51~(0.08) & 13.65~(0.09) \cr
\noalign{\vskip 6pt}
\noalign{\hrule}
\noalign{\vskip 6pt}
NY93 &13.58&\cite{ny93}& $0-350$  & 16 & 13.04~(0.09) & 13.31~(0.09) \cr
BONN &14.40&\cite{bonn}& $0-325$  & 15 & 13.72~(0.10) & 13.74~(0.10) \cr
PARIS&14.43&\cite{paris}&$0-350$  & 16 & 13.46~(0.08) & 13.76~(0.09) \cr
\noalign{\vskip 10pt}}
\hrule}}
\vfill
\eject
Table~III. Table of optimal $g^2/4\pi$ and $\chi^2$ values obtained from
$\chi^2$ maps. Solution A has only the one-pion exchange (OPE) contribution
for J$>$6. Solution B has only the OPE contribution for J$>$5. Solution C
is the same as A, with a plane-wave Born approximation for the OPE.
The region below 400 MeV contains 2170 pp and 3532 np data. The region
below 1600 MeV contains 12838 pp and 11171 np data.
\vskip 10pt
\centerline{
\vbox{\offinterlineskip
\hrule
\hrule
\halign{\hfill#\hfill&\qquad\hfill#\hfill&\qquad\hfill#\hfill&
\qquad\hfill#\hfill&\qquad\hfill#\hfill&\qquad\hfill#\hfill&
\qquad\hfill#\hfill\cr
\noalign{\vskip 6pt}
Solution&$g^2/4\pi$&$\chi^2$&$g^2/4\pi$&
$\chi^2$&$g^2/4\pi$&$\chi^2$ \cr
\hbox{\hskip .1cm}&(pp data only)&\hbox{\hskip .1cm}&(np data only)&
\hbox{\hskip .1cm}&(pp and np data)&\hbox{\hskip .1cm}\cr
\noalign{\vskip 6pt}
\noalign{\hrule}
\noalign{\vskip 6pt}
A(0-400~MeV)&13.61(0.09)& 3034 & 14.16(0.12) & 4533 & 13.80(0.07) & 7582 \cr
\noalign{\vskip 6pt}
B(0-400~MeV)&13.70(0.08)& 3055 & 14.07(0.11) & 4555 & 13.83(0.07) & 7617 \cr
\noalign{\vskip 6pt}
C(0-400~MeV)&14.42(0.08)& 3046 & 14.51(0.11) & 4554 & 14.38(0.06) & 7602 \cr
\noalign{\vskip 6pt}
(0-1600~MeV)&13.67(0.06)&22030 & 13.42(0.04) & 17625& 13.51(0.03) & 39668\cr
\noalign{\vskip 10pt}}
\hrule}}
\vfill
\eject

\end{document}